\documentstyle[a4,12pt]{article}
\begin{document}
\pagestyle{empty}
\begin{titlepage}
\rightline{IC/96/175}
\rightline{UTS-DFT-96-01}
\vspace{2.0 truecm}
\begin{center}
\begin{Large}
{\bf Constraints on $Z^\prime$ from $W^+W^-$ production}\\[0.3cm]
{\bf at the NLC with polarized beams}
\end{Large}

\vspace{2.0cm}

{\large  A.A. Pankov\hskip 2pt\footnote{Permanent address: Gomel
Polytechnical Institute, Gomel, 246746 Belarus.\\ 
E-mail PANKOV@GPI.GOMEL.BY}
}\\[0.3cm]
International Centre for Theoretical Physics, Trieste, Italy\\
Istituto Nazionale di Fisica Nucleare, Sezione di Trieste, 34127 Trieste,
Italy

\vspace{5mm}

{\large  N. Paver\hskip 2pt\footnote{Also supported by the Italian
Ministry of University, Scientific Research and Technology (MURST).}
}\\[0.3cm]
Dipartimento di Fisica Teorica, Universit\`{a} di Trieste, 34100
Trieste, Italy\\
Istituto Nazionale di Fisica Nucleare, Sezione di Trieste, 34127 Trieste,
Italy\\
\end{center}

\vspace{2.0cm}

\begin{abstract}
\noindent 
We discuss the potential of NLC500 and NLC1000 to probe $Z$-$Z^\prime$ mixing 
and mass by the reaction $e^+e^-\to W^+W^-$ with longitudinally polarized 
electrons. We perform a model-independent analysis of the deviations from the 
Standard Model, and apply it to a specific class of extended weak gauge models.
Results indicate that the corresponding bounds at the NLC500 complement the 
present ones obtained from LEP1, and rapidly become quite stringent at 
the higher energy of the NLC1000. Also, we emphasize the importance of 
initial beam polarization in improving the sensitivity to mixing.  
\par 

\vspace*{3.0mm}

\noindent  
\end{abstract}
\end{titlepage}

\pagestyle{plain}
\setlength{\baselineskip}{1.3\baselineskip}
\section{Introduction}
A rather common feature of extended electroweak models is the 
prediction that one (or more) neutral heavy gauge bosons $Z^\prime$ should 
exist. The mass of such objects is largely unknown, as it cannot be 
theoretically estimated by a dynamical calculation, but it might be 
in the TeV range and thus, hopefully, in the reach  
of future higher energy machines. Clearly, the knowledge of $Z^\prime$ 
parameters such as $M_{Z^\prime}$, the $Z^\prime$ couplings to fermions 
and its mixing angle with the standard $Z$, is essential to test extended 
theories, and in particular their gauge and Higgs structures. 
Many attempts have been made to phenomenologically 
derive indications on the $Z^\prime$ properties from present data, and to 
develop strategies to search for the manifestations of such particles in the 
data from next-generation high energy machines. \par 
At present, direct $Z^\prime$ production searches at the $p$-$\bar p$ 
Tevatron collider indicate a lower limit on $M_{Z^\prime}$ of the order of 
$500\hskip 2pt GeV$, which means that almost certainly the LEP200 will be 
below the threshold for directly studying the $Z^\prime$ peak, while the 
the planned next-linear $e^+e^-$ colliders might be near the discovery level. 
If the CM energy were still not high enough to 
produce such heavy particle, $Z^\prime$ searches would focus on 
possible `indirect' manifestations through deviations of observables from 
the Standard Model (SM) predictions. If such deviations were not found within 
the expected accuracy, then constraints on the various parameters could be 
derived. More optimistically, 
in the event that some deviation was observed, in principle such information 
could be used to shed light on the properties of the $Z^\prime$. Of course, 
a related problem would be to have some criterion to distinguish that 
effect from alternative sources of nonstandard physics, potentially 
also contributing deviations from the SM. 
\par   
Studies of the annihilation $e^-e^+\to f\bar f$ at LEP1 have lead to 
restrictions on the ${Z-Z^\prime}$ mixing angle and the $Z^\prime$ mass 
comparable with the direct searches at the Tevatron and, in the perspective 
both of LEP200 and of the next-linear-collider projects, the sensitivity of 
such reaction to the $Z^\prime$ has been extensively analysed, also recently 
\cite{chiappetta}-\cite{verzegnassi}.\par 
With the increased $e^+e^-$ energy available at these machines, also 
the reaction 
\begin{equation} e^++e^-\to W^+W^-\label{proc}\end{equation}
should represent a convenient tool to search for $Z^\prime$ 
effects \cite{pankov}. Indeed, in this process, lack of gauge cancellation 
among the different amplitudes due to nonstandard physics should lead to 
deviations from the SM cross section rapidly increasing with energy and 
therefore, in principle, to enhanced sensitivity to the existence of the 
$Z^\prime$ if efficient $W^+W^-$ reconstruction could be performed. 
Moreover, it turns out that the strongest sensitivity of process (\ref{proc}) 
to nonstandard effects would be obtained if initial beams 
were longitudinally polarized with both kinds of polarization available, 
so that the information from the separately measurable cross sections for 
left-handed and right-handed electrons could be combined. On the one hand, 
that would lead to stringent restrictions on the ${Z-Z^\prime}$ mixing 
angle. On the other hand, model-independent separate bounds on the 
anomalous ($WW\gamma$) and ($WWZ$) couplings could be derived 
in case the deviation from the SM was attributed to this kind of 
source \cite{andreev}. In this 
note, we will discuss manifestations of the $Z^\prime$ in $e^+e^-\to W^+W^-$ 
at future $e^+e^-$ colliders taking into account, in addition to propagator 
effects as in  Ref.~\cite{verzegnassi}, also the ${Z-Z^\prime}$ mixing. 
Starting from a model-independent parametrization of the deviations from the 
SM amplitude, we derive bounds on nonstandard couplings and then 
apply such 
constraints to the class of extended electroweak models based on $E_6$ gauge 
symmetry. The high sensitivity of process (\ref{proc}) to the parameters of 
such extended models, namely the ${Z-Z^\prime}$ mixing angle and the heavier 
neutral gauge boson mass, will be shown. In particular, the essential role of 
initial beams polarization in this kind of analyses will be emphasized.

\section{Model independent bounds} 
The SM amplitude for process (\ref{proc}), in Born approximation, is divided 
into $t$-channel amplitude (originating from neutrino exchange) and $s$-channel 
amplitude (induced by photon and $Z$ exchange):  
${\cal M}={\cal M}_t+{\cal M}_s$. With $\lambda(=-\lambda^\prime)=
\pm 1/2$ the electron (positron) helicity, the $t$-channel amplitude has the 
form
\begin{equation}{\cal M}^\lambda_t=\frac{2\lambda-1}{4ts_W^2}\times 
{\cal T}^\lambda(s,\theta),\label{amplit}\end{equation}  
where $\sqrt s$ and $\theta$ are the total energy and the $W^-$ 
production angle in the CM frame, $t=M_W^2-s(1-\beta_W\cos\theta)/2$ with 
$\beta_W=\sqrt{1-4M_W^2/s}$, and the explicit form of the kinematical 
coefficient ${\cal T}^\lambda(s,\theta)$ is not essential for our discussion 
and is omitted for simplicity, as it can easily be found in the 
literature \cite{gounaris}.\par
The $s$-channel amplitude is given by
\begin{equation} 
{\cal M}_s^\lambda=\left(-\frac{g_{WW\gamma}}{s}
+\frac{g_{WWZ}(v-2\lambda a)}{s-M^2_Z}\right)
\times{\cal G}^\lambda(s,\theta)\label{amplis}\end{equation}
and, as in the case of Eq.~(\ref{amplit}), we refer to \cite{gounaris} 
for the expression of the kinematical coefficient 
${\cal G}^\lambda(s,\theta)$.\par 
The SM values of the couplings in Eq.~(\ref{amplis}) are: $g_{WW\gamma}=1$ and 
$g_{WWZ}=\cot\theta_W$ with $\theta_W$ the electroweak mixing angle 
($\sin^2\theta_W\simeq 0.231$); $v=(T_{3,e}-2Q_e\hskip 2pt s_W^2)/2s_Wc_W$ 
and $a=T_{3,e}/2s_Wc_W$ with $T_{3,e}=-1/2$ ($s_W=\sin\theta_W$, 
$c_W=\cos\theta_W$); $M_Z^2= M_W^2/c_W^2$ at the leading order.\par 
In the extended electroweak models considered here, the neutral vector 
boson sector consists, in addition to the photon, of a light $Z_1$ (with 
mass $M_{Z_1}\approx M_Z$, to be identified to the $Z$ in the limit of the 
SM), and a heavy $Z_2$ (with, expectedly, $M_{Z_2}$ much greater than $M_Z$). 
Both exchanges contribute to the $s$-channel amplitude, which now reads:
\begin{equation}{\cal M}_s^\lambda=\left(-\frac{g_{WW\gamma}}{s}
+\frac{g_{WWZ_1}(v_1-2\lambda a_1)}{s-M_{Z_1}^2}
+\frac{g_{WWZ_2}(v_2-2\lambda a_2)}{s-M_{Z_2}^2}\right)
\times{\cal G}^\lambda(s,\theta).\label{amplis1}\end{equation}
Here, $Z_1$ and $Z_2$ denote `physical' mass-eigenstates; $g_{WWZ_1}$ and 
$g_{WWZ_2}$ are the corresponding couplings to $W^+W^-$, both assumed of 
the usual Yang-Mills form, and 
$(v_1,\hskip 1pt a_1)$ and $(v_2,\hskip 1pt a_2)$ are, respectively, the  
vector and axial-vector couplings to $e^+e^-$. Clearly, the values of these 
couplings depend on the extended model under consideration. In the sequel, 
the notation $Z$ and $Z^\prime$ will indicate the Standard Model 
$Z$-particle and the heavy neutral boson weak gauge-eigenstate. \par 
Concerning the $Z_1$ couplings to electrons, present constraints from 
experimental data \cite{langacker} indicate that their values should be 
rather close to the SM values $v$ and $a$ listed above. Thus, the deviations 
$\Delta v=v_1-v$ and $\Delta a=a_1-a$ should be small numbers, to be treated 
as a perturbation, and the same is true for the mass-shift 
$\Delta M=M_{Z}-M_{Z_1}$, with $\Delta M>0$ if this is due to 
${Z-Z^\prime}$ mixing.\par 
In linear approximation, as justified by the expected smallness of 
these nonstandard effects, all deviations can be conveniently parametrized 
so as to rewrite the $s$-channel amplitude (\ref{amplis1}) in the same form 
as the SM one:
\begin{equation} 
{\cal M}_s^\lambda=\left(-\frac{g^*_{WW\gamma}}{s}
+\frac{g^*_{WWZ}(v-2\lambda a)}{s-M^2_Z}\right)
\times{\cal G}^\lambda(s,\theta),\label{amplis2}\end{equation}
where the `effective' gauge boson couplings $g^*_{WW\gamma}$ and $g^*_{WWZ}$ 
are defined as: 
\begin{equation} 
g^*_{WW\gamma}\equiv 1+\delta_\gamma\equiv 
1+\delta_\gamma(Z_1)+\delta_\gamma(Z_2),\label{deltagamma}\end{equation}
\begin{equation}
g^*_{WWZ}\equiv\cot\theta_W+\delta_Z\equiv \cot\theta_W
+\delta_Z(Z_1)+\delta_Z(Z_2),
\label{deltaz}\end{equation}
with
\begin{equation}\delta_\gamma(Z_1)=v\hskip 1pt g_{WWZ_1}
\left(\frac{\Delta a}{a}-\frac{\Delta v}{v}\right)
\left(1+\Delta\chi\right)\hskip 1pt\chi;\ \ \ \  
\delta_\gamma(Z_2)=v\hskip 1pt g_{WWZ_2}\hskip 1pt
\left(\frac{a_2}{a}-\frac{v_2}{v}\right)\hskip 1pt \chi_2,
\label{coupl1}
\end{equation}
\begin{equation}
\delta_Z(Z_1)=-\cot\theta_W+g_{WWZ_1}
\left(1+\frac{\Delta a}{a}\right)\left(1+\Delta\chi\right);
\qquad 
\delta_Z(Z_2)=
g_{WWZ_2}\hskip 2pt \frac{a_2}{a}\hskip 2pt \frac{\chi_2}
{\chi}.\label{coupl2}\end{equation}
In Eqs.~(\ref{coupl1}) and (\ref{coupl2}), neglecting the gauge boson widths:
\begin{equation}
\chi (s)=\frac{s}{s-M^2_Z};\qquad \chi_2 (s)=\frac{s}{s-M^2_{Z_2}};
\qquad \Delta\chi (s)=-\frac{2M_Z\Delta M}{s-M^2_Z}.\label{chi}
\end{equation}
As it will be emphasized in the sequel, this general parametrization is rather 
useful for phenomenological purposes, in order to discuss the deviations 
from the SM in the context of different classes of extended models 
contributing to the deviations (\ref{coupl1}) and (\ref{coupl2}). \par  
As indicated by the notations, in Eqs.~(\ref{deltagamma}) and (\ref{deltaz}) 
$\delta_\gamma(Z_1)$ and $\delta_Z(Z_1)$ originate from modifications of the 
$Z$ couplings plus the, generally possible, $Z$ mass-shift from the SM value 
accounted by $\Delta M$ in (\ref{chi}). Instead, $\delta_\gamma(Z_2)$ 
and $\delta_Z(Z_2)$ represent the `direct' contribution of the $Z_2$. 
One can notice that the resulting expressions for $g^*_{WW\gamma}$ and 
$g^*_{WWZ}$ coincide with those used in \cite{verzegnassi} in the limit of 
retaining only the structure corresponding to $\delta_\gamma(Z_2)$ and 
$\delta_Z(Z_2)$. From the general form of Eqs.~(\ref{amplis2})-(\ref{deltaz}), 
nonvanishing values of $\delta_\gamma$ and $\delta_Z$ can also occur as 
the consequence of anomalous trilinear gauge boson couplings. Actually, in 
effective theories of nonstandard anomalous trilinear gauge boson couplings 
\cite{buchmuller, zeppenfeld, gavela}, at the leading dimension 
$\delta_\gamma=0$.\footnote{The problem of distinguishing this alternative 
source of nonstandard effects from mixing will be discussed in a separate 
paper.}  
\par  
Assuming that the new physics does not involve the charged current couplings 
$(We\nu)$, which then retain their SM values, the $t$-channel amplitude 
remains identical to Eq.~(\ref{amplit}). Consequently, the differential cross 
section of process (\ref{proc}) at a given CM energy will have the same form 
as in the SM, except that in general its values will numerically differ from 
the SM predictions due to the nonstandard effects introduced above. 
Accordingly, we can represent such effect by the relative deviation of the 
cross section (either differential or integrated in some angular range) from 
the SM prediction: 
\begin{equation}\Delta\equiv\frac{\Delta\sigma}{\sigma_{SM}}=
\frac{\sigma-\sigma_{SM}}{\sigma_{SM}},\label{delta}\end{equation} 
which brings information on the free, independent, parameters 
$\delta_\gamma$ and $\delta_Z$ in Eqs.~(\ref{deltagamma}) and (\ref{deltaz}). 
\par 
If a nonvanishing value of $\Delta$ was experimentally measured at some level 
of accuracy, the values of such parameters could be determined and possibly 
used to learn about the properties of the related nonstandard physics. 
Alternatively, in the case of no observation, one could derive numerical 
bounds on $\delta_\gamma$ and $\delta_Z$, and therefore constrain the various 
extended models, at some confidence level that depends on the attainable 
sensitivity of the experiment. In this regard, assuming small deviations, 
$\Delta$ is expressed as a linear combination of $\delta_\gamma$ and 
$\delta_Z$ with coefficients which, generally, increase with $s$. Conversely, 
the SM cross section decreases as $1/s$ (at least) due to the gauge 
cancellation among the various amplitudes. Therefore, if we parametrize the 
sensitivity of process (\ref{proc}) to $\delta_\gamma$ and $\delta_Z$ by, 
e.g., the ratio ${\cal S}=\Delta/(\delta\sigma/\sigma)$ with 
$\delta\sigma/\sigma$ the attainable statistical uncertainty on the SM cross 
section, such sensitivity is expected to increase with energy, even at fixed 
integrated luminosity (basically, as ${\cal S}\propto\sqrt{L_{int}s}$).\par 
As discussed previously \cite{andreev}, initial electron beam longitudinal 
polarization, and the related possibility to individually measure the cross 
sections for $e^-_Le^+$ and $e^-_Re^+$ ($\sigma^-$ and $\sigma^+$), would 
substantially improve the sensitivity to the couplings $\delta_\gamma$ and 
$\delta_Z$. In this regard, the measurement of $\sigma^+$ would be of 
particular interest in two respects. Firstly, although giving a much lower 
statistics as being suppressed by $\gamma$-$Z$ gauge compensation, it is most 
sensitive to the nonstandard effects considered here because it is free of the 
unmodified $t$-channel amplitude which numerically dominates 
$\sigma^-$ and $\sigma^{unpol}$ as well 
($\sigma^{unpol}\approx (1/2)\sigma^-$). Secondly, by specifying $\lambda$, 
Eq.~(\ref{amplis2}) directly shows that 
\begin{equation}
\Delta\sigma^-\propto \delta_\gamma-\delta_Z\cdot g^L_e\chi; \qquad 
\Delta\sigma^+\propto\delta_\gamma-\delta_Z\cdot g^R_e\chi, 
\label{correl}\end{equation}
where $g^R_e=v-a=\tan\theta_W\simeq 0.55$ and 
$g^L_e=v+a=g^R_e\left(1-1/2s_W^2\right)\simeq -0.64$. Thus, by themselves, 
$\sigma^-$ (or $\sigma^{unpol}$) and $\sigma^+$ only provide correlations 
among $\delta_\gamma$ and $\delta_Z$, rather than true limits. These 
correlations are represented as bands in the ${\delta_\gamma-\delta_Z}$ plane 
of Fig.~1, with a width proportional to the corresponding sensitivities, 
and a relative angle of approximately 60 degrees. The figure clearly 
illustrates that, in contrast with the case where only the unpolarized cross 
section is measured, and therefore in principle an `unlimited' area is allowed 
to $\delta_\gamma$ and $\delta_Z$,  
the combination of left-handed and right-handed cross sections is essential 
to obtain a finite allowed region by the intersection of the corresponding 
bands.\par 
Numerically, Fig.~1 refers to the NLC energy $\sqrt s=500\hskip 2pt GeV$, 
assuming $L_{int}=50\hskip 2pt fb^{-1}$, and to cross sections integrated 
over the range $\vert\cos\theta\vert\leq 0.98$. Concerning polarization, in 
practice the degree of initial electron longitudinal polarization $P_L$ will 
be quite near, but not be exactly equal, to unity. Therefore, the measured 
cross section will be a linear combination of purely polarized cross 
sections \cite{andreev}:
\begin{equation}
\frac{d\sigma}{d\cos\theta}=\frac{1}{4}\left[\left(1+P_L\right)
\frac{d\sigma^+}{d\cos\theta}+\left(1-P_L\right)
\frac{d\sigma^-}{d\cos\theta}\right].\label{longi}\end{equation}
In Fig.~1, the notation $\sigma^R$ and $\sigma^L$ refers to the values 
$P_L=0.9$ and $P_L=-0.9$, respectively. Such values of $P_L$ seem to be 
obtainable at the NLC \cite{prescott}.\par
The sensitivity of $\sigma^L$ and $\sigma^R$ to $\delta_\gamma$ and $\delta_Z$ 
has been assessed numerically by dividing the considered angular range into 
10 equal `bins', and defining a $\chi^2$ function in terms of the expected 
number of events $N(i)$ in each bin:
\begin{equation}
\chi^{2}=\sum^{bins}_i\left[\frac{N_{SM}(i)-N(i)}
{\delta N_{SM}(i)}\right]^2,\label{chi2}\end{equation}
where the uncertainty on the number of events $\delta N_{SM}(i)$ combines 
both statistical and systematic errors as
\begin{equation} \delta N_{SM}(i)=
\sqrt{N_{SM}(i)+\left(\delta_{syst}N_{SM}(i)\right)^2},
\label{deltan} \end{equation}
(we assume $\delta_{syst}=2\%$).
In Eq.~(\ref{chi2}), $N(i)=L_{int}\sigma_i\varepsilon_W$, with 
\begin{equation}
\sigma_i\equiv\sigma(z_i,z_{i+1})=
\int \limits_{z_i}^{z_{i+1}}\left({d\sigma}\over{dz}\right)dz,
\label{sigmai}\end{equation}
and $z=\cos\theta$. Also, $\varepsilon_W$ is the efficiency for $W^+W^-$ 
reconstruction, for which we take the channel of lepton pairs ($e\nu+\mu\nu$) 
plus two hadronic jets, giving $\varepsilon_W\simeq 0.3$ from the relevant 
branching ratios. An analogous procedure is followed to evaluate $N_{SM}(i)$.
\par 
As a criterion to derive allowed regions for the deviations of the coupling 
constants in the case no deviations from the SM were observed, and in this 
way to assess the sensitivity of process (\ref{proc}) to $\delta_\gamma$ and 
$\delta_Z$, we impose that $\chi^2\leq\chi^2_{crit}$, where $\chi^2_{crit}$ 
is a number that specifies the chosen confidence level. With two independent 
parameters in Eq.~(\ref{correl}), the $95\%$ CL is obtained by choosing 
$\chi^2_{crit}=6$ \cite{pdg}.
\par 
From the numerical procedure outlined above, we obtain the bands 
allowed to $\delta_\gamma$ and $\delta_Z$ by the polarized cross sections 
$\sigma^R$ and $\sigma^L$ (as well as $\sigma^{unpol}$) depicted in Fig.~1. 
This figure manifestly shows the significant role of the combination of 
polarized measurements in restricting the limits by the intersection of the 
two bands.

\section{Bounds on extended models and conclusions}
In this section, we apply the information on $\delta_\gamma$ and $\delta_Z$ 
given in Fig.~1 to the tests of extended models where the gauge group contains 
at least one additional $U(1)^\prime$ factor, therefore one new neutral gauge 
boson $Z^\prime$  \cite{rizzo}. Specifically, we focus on the `conventional' 
class of models with an $E_6$ origin (either inspired by superstring theories 
or not), where the Higgs fields transform either as doublets or singlets of 
$SU(2)_L$. 
\par
In general, the neutral gauge boson mass matrix in the basis of weak gauge 
eigenstates $Z$ and $Z^\prime$ takes the form
\begin{equation}M^2=\pmatrix{M_Z^2&\delta M^2\cr \delta M^2&M_{Z^\prime}^2}.
\label{matrix}\end{equation}
Diagonalization of (\ref{matrix}) leads to the mass eigenstates $Z_1$ and 
$Z_2$ {\it via} the rotation 
\begin{eqnarray} Z_1&=&Z\cos\phi+Z'\sin\phi\nonumber \\
Z_2&=&-Z\sin\phi+Z'\cos\phi, \label{mixing}\end{eqnarray}
where, by convention, $M_{Z_1}<M_{Z_2}$ and $\phi$ is the mixing angle:
\begin{equation}\tan^2\phi=\frac{M^2_Z-M^2_{Z_1}}{M^2_{Z_2}-M^2_Z}
\simeq\frac{2M_Z\Delta M}{M^2_{Z_2}}.\label{phi}\end{equation}
\par
The mixing of the SM $Z$ with the heavier $Z^\prime$ leads to an `observed'  
mass of the lighter neutral gauge boson $Z_1$ shifted from $M_Z$ by the 
positive mass-shift $\Delta M$ introduced in Eqs.~(\ref{chi}) and (\ref{phi}). 
Basically, model-independent information on $\Delta M$ can be derived, 
for example along the lines of Ref.~\cite{barger}, from the radiatively 
corrected value of $M_Z$ evaluated in the SM. The starting point is the 
relation
\begin{equation}
M_W^2=\frac{A}{s_W^2 (1-\Delta r)},
\label{mw}\end{equation}
where $A=\pi\alpha(m_e)/\sqrt 2 G_F$ and $\Delta r$ takes into account 
radiative corrections (depending on $M_W$, $M_Z$, $m_{\it top}$ and the 
Higgs mass $m_H$).\footnote{In extended electroweak models, $m_H$ effectively 
indicates the combined contribution of scalar fields to the radiative 
corrections to $M_W$.} Moreover, in the on-shell renormalization 
scheme \cite{sirlin1, sirlin2}, the electroweak mixing angle including 
radiative corrections is expressed as 
\begin{equation} 
s_W^2=1-M_W^2/M_Z^2, \label{sw}\end{equation}
reflecting the weak isospin doublet or singlet character of the Higgs fields.   
The fact that the $W$ mass and couplings are unaffected (at tree level) by 
the $Z^\prime$ allows the use of the $M_W$ measured at the Tevatron to extract 
$s_W^2$ in terms of $M_W$, $m_{\it top}$ and $m_H$ from Eq.~(\ref{mw}). 
Replacing the so determined $s_W$ into Eq.~(\ref{sw}) determines the 
predicted SM value of $M_Z$ in terms of $m_{\it top}$ and $m_H$, which must 
be compared to the value of $M_{Z_1}$ determined by LEP data. From 
the current values of $M_W$, $M_{Z_1}$ and $m_{\it top}$ \cite{pdg}, and for 
$m_H$ in the range $100$-$500\hskip 2pt GeV$, one finds for $\Delta M$ an 
upper limit of the order of $200\hskip 2pt MeV$. This is compatible with the 
updated analysis of $\Delta M$ in Ref.~\cite{altarelli}. 
Furthermore, current limits on the mixing angle $\vert\phi\vert $ are in the 
range $10^{-3}-10^{-2}$, mostly from LEP data 
\cite{langacker, altarelli, aleph}. 
\par 
In addition to the $Z$ mass-shift, the $Z$-$Z^\prime$ mixing induces a change 
in the couplings of the $Z$ to fermions. In the considered models, the neutral 
current coupled to the $Z^\prime$ is parametrized in terms of an angle 
$\beta$ specifying the orientation of the $U(1)^\prime$ generator in the $E_6$ 
group space \cite{zwirner}. Explicitly: 
\begin{equation}
v^\prime=\frac{\cos\beta}{c_W\hskip 1pt\sqrt6};\qquad\quad 
a^\prime=\frac{1}{2\hskip 1pt c_W\sqrt 6}\left(\cos\beta+
\sqrt{\frac{5}{3}}\sin\beta\right).\label{vprime}\end{equation}
In general, $\cos\beta$ can range from $-1$ to $+1$. Special values are 
$\beta=0;\hskip 2pt \pi/2;\hskip 2pt \pi-\arctan\sqrt{5/3}\approx 128^\circ$, 
which specify the so-called $\chi$, $\psi$ and $\eta$ models, respectively.
\par 
While being in general an independent parameter, in specific $E_6$ models 
where the $SU(2)_L$ doublet and singlet Higgses all arise from the {\bf 27} 
representation of $E_6$, the mixing angle $\phi$ can be related to the values 
of $M_{Z_1}$ and $M_{Z_2}$ \cite{rizzo}. For $M_{Z_2}$ much larger than 
$M_{Z_1}$ the relation can be written to a good approximation as:  
\begin{equation}\phi\simeq -\sin^2\theta_W\ {\sum_{i}<\Phi_i>^2
I^i_{3L}Q^{\prime}_i
\over\sum_{i}<\Phi_i>^2(I^i_{3L})^2}={\cal C}\
{{\displaystyle M^2_1}\over{\displaystyle M^2_2}}\label{models}\end{equation}
where, depending on the model, $<\Phi_i>$ are the Higgs vacuum expectation 
values spontaneously breaking the gauge symmetry, and $Q^{\prime}_i$ are their 
$U(1)^{\prime}$ charges. For example, in the case of $E_6$ `superstring' 
inspired models, ${\cal C}$ can be expressed as 
\begin{equation}{\cal C}=4s_W\left(\frac{\cos\beta}{2\sqrt 6}-
{\sigma-1\over\sigma+1}\hskip 2pt\frac{\sqrt10\hskip 1pt\sin\beta}{12}\right),
\label{c}\end{equation}
where $\sigma$ is a ratio of vacuum expectation values squared.
\par 
With reference to the couplings introduced in Eq.~(\ref{amplis1}), 
taking Eq.~(\ref{phi}) into account we have to first order in $\phi$, in a 
self-explaining notation:
\begin{equation}
(v_1,\hskip 1pt a_1)\simeq (v+v^\prime\hskip 1pt\phi,\hskip 1pt 
a+a^\prime\hskip 1pt\phi)
\Rightarrow (\Delta v,\hskip 1pt\Delta a)\simeq (v^\prime\hskip 1pt \phi, 
a^\prime\hskip 1pt\phi),\label{v1}\end{equation} 
\begin{equation}
(v_2,\hskip 1pt a_2)\simeq(-v\hskip 1pt\phi+v^\prime,\hskip 1pt 
-a\hskip 1pt\phi+a^\prime),\label{v2}\end{equation}
and
\begin{equation}
g_{WWZ_1}\simeq g_{WWZ}; \qquad\quad g_{WWZ_2}\simeq -g_{WWZ}\hskip 1pt\phi.
\label{g}\end{equation}
Replacing Eqs.~(\ref{v1})-(\ref{g}) into (\ref{coupl1}) and (\ref{coupl2}), 
one finds the general form of $\delta_\gamma$ and $\delta_Z$ for $E_6$ 
models:
\begin{equation}
\delta_\gamma=v\hskip 1pt\cot\theta_W\hskip 1pt\phi\hskip 1pt 
\left(\frac{a^\prime}{a}-\frac{v^\prime}{v}\right)
\left(1-\frac{\chi_2}{\chi}+\Delta\chi\right)\chi,
\label{ddeltag}\end{equation}
\begin{equation}
\delta_Z=\cot\theta_W\hskip 1pt\left[\phi
\frac{a^\prime}{a}\left(1-\frac{\chi_2}{\chi}\right)+
\Delta\chi\right].\label{ddeltaz}
\end{equation}
Although present in general, at the order we are working here we could safely 
neglect $\Delta\chi$ which is of order $\phi^2$ due to (\ref{chi}) and 
(\ref{phi}).\footnote{Taking $\Delta\chi$ into account would slightly shift 
the origin in Fig.~1.} In this case, from Eqs.~(\ref{ddeltag}) and 
(\ref{ddeltaz}), there is the linear relation between $\delta\gamma$ and 
$\delta_Z$  which only depends on the $Z^\prime$ couplings to fermions and is 
independent from $\phi$ and $M_{Z_2}$:
\begin{equation}
\delta_Z=\delta_\gamma\hskip 1pt\frac{1}{v\hskip 1pt\chi}\hskip 1pt
\frac{(a^\prime/a)}{(a^\prime/a)-(v^\prime/v)}.\label{relation}\end{equation} 
\par 
Eq.~(\ref{relation}) represents straight lines in the plane of Fig.~1, and 
we explicitly report the representatives of models $\psi$, $\chi$ 
and $\eta$. As one can see, sensitivities are different: as indicated by the 
intersection points of these lines with the allowed bands, 
while $\sigma^R$ mostly constrains the models $\psi$ and $\eta$, $\sigma^L$ is 
the most sensitive to the $\chi$ model. Furthermore, the allowed range of 
variation of $\delta_\gamma$ and $\delta_Z$ for the specific models is defined 
by the intersections of the corresponding lines with the boundaries of the, 
model-independent, allowed region obtained from the combination of $\sigma^R$ 
and $\sigma^L$. One should also remark that, without the information 
from $\sigma^R$, models represented by lines almost parallel to the 
band determined by $\sigma^{unpol}$ (e.g., the $\psi$ model), would remain 
essentially unconstrained.
\par
The ranges of $\delta_\gamma$ and $\delta_Z$ allowed to the specific models 
in Fig.~1 can be translated into limits on the mixing angle $\phi$ and the 
heavier gauge boson mass $M_{Z_2}$, using 
Eqs.~(\ref{ddeltag})-(\ref{relation}). Starting our discussion from the 
$\psi$ model, the resulting allowed region (at the 95\% CL) in the 
($\phi,M_{Z_2}$) plane is limited in this case by the thick solid line in 
Fig.~2. We have chosen $\Delta M$ equal to the previously mentioned upper 
limit of $200\hskip 2pt MeV$, although the limiting curves do not appreciably 
depend on the specific value of this quantity. Also, the indicative current 
bound on $M_{Z_2}$ from direct searches is reported in this figure.
\par 
Fig.~2 shows that the process $e^+e^-\to W^+W^-$ at $500\hskip 2pt GeV$ 
has a potential sensitivity to the mixing angle $\phi$ of the order of 
$10^{-4}-10^{-3}$, depending on the mass $M_{Z_2}\gg M_{Z_1}$ ranging up from 
the current lower bound of $500\hskip 2pt GeV$. Specifically, as it is 
seen from Eqs.~(\ref{ddeltag}) and (\ref{ddeltaz}), for the higher masses 
$M_{Z_2}$ much larger than $\sqrt{2 s}$ such that the $Z_2$ exchange 
contribution $\vert\chi_2/\chi\vert$ is much less than unity, the limiting 
contour is mostly determined by the modification (\ref{v1}) of the $Z$ 
couplings to electrons. Asymptotically, in the limit $M_{Z_2}\to\infty$ where   
$\left(1-\chi_2/\chi\right)\to 1$, the limiting angle is 
$\phi\sim 10^{-3}$. In the region $\sqrt s< M_{Z_2}<\sqrt{2s}$ 
one has to account for (at least) the imaginary part of the $Z_2$ propagator 
$\chi_2$ by the replacement $M_{Z_2}^2\to M_{Z_2}^2-iM_{Z_2}\Gamma_{Z_2}$ in 
Eq.~(\ref{chi}). Clearly, the sensitivity to $\phi$ in this mass range is 
dominated by the interference $\chi_2/\chi$ in Eqs.~(\ref{ddeltag}) and 
(\ref{ddeltaz}) and is 
enhanced with respect to that expected in the higher $Z_2$ mass range by a 
factor which, for the specific values ${\displaystyle M_{Z_2}=\sqrt s\pm 
\frac{\Gamma_{Z_2}}{2}}$, can reach the value 
\begin{equation}  
\vert{\rm Re}\frac{\chi_2}{\chi}\vert\simeq
\frac{M_{Z_2}}{2\Gamma_{Z_2}}\simeq 20.\label{enhancement}\end{equation}     
The factor 20 in Eq.~(\ref{enhancement}) conservatively assumes 
$n_g=3$ exotic heavy fermions generations that, in addition to the 
conventional fermions, can contribute to $Z_2$ decay without significant 
phase space suppression. In this case, one approximately expects 
$\Gamma_{Z_2}\simeq 0.025 M_{Z_2}$, independent of $\cos\beta$ and $\phi$ 
\cite{barger1}. In this situation, from the above factor we 
can qualitatively estimate a sensitivity to values of $\phi$ of the order of 
$10^{-4}$ or less. Even more stringent numerical constraints, by a factor 
of order 2-5, would be obtained if $n_g=0$, which would imply a smaller value 
of $\Gamma_{Z_2}$.
\par 
To complete the discussion on the bounds from process (\ref{proc}), one should 
make a comparison of the results presented so far with the maximal allowed 
region to $\phi$ and $M_{Z_2}$ determined in a model- and process-independent 
way by the limit on the off-diagonal mass-matrix element. The relevant 
boundary contours, Eq.~(\ref{phi}), are represented in Fig.~2 by the dotted 
lines, corresponding to the chosen upper limit for $\Delta M$. Clearly, for 
values of $M_{Z_2}$ higher than the intersection of these lines with the thick 
solid line, Eq.~(\ref{phi}) gives the most stringent bounds on 
($\phi,\hskip 2pt M_{Z_2}$). In this regard, it is useful to compare also with 
the specific `superstring inspired' model previously introduced, where 
$\phi$ and $M_{Z_2}$ are uniquely related through Eq.~(\ref{models}). Such 
relation is illustrated in Fig.~2 by the continuous thin lines, which 
correspond to $\sigma=0$ and $\sigma=\infty$ in (\ref{c}) respectively, as 
representative cases. \par
In Fig.~2, we consider also the limits on $\phi$ and $M_{Z_2}$ that are 
expected from the annihilation into lepton pairs at 
$\sqrt s=0.5\hskip 2pt TeV$. It should be 
emphasized that this process can give alternative (and competitive) bounds, 
through the combination of the (almost $\phi$-independent) lower bound on 
$M_{Z_2}$ with the area allowed by Eq.~(\ref{phi}).\footnote{For this process, 
we use a systematic uncertainty $\delta_{syst}=1\%$.}
\par 
Up to this point, the discussion has been based on the assumption that signals 
of the $Z_2$ were not observed within the accuracy, and the corresponding 
limits have been assessed. One could consider the reverse scenario, and 
assume that the $Z_2$ is observed with a mass lower than the discovery limit 
in Fig.~2. Clearly, from higher peak cross sections and almost 
$\phi$-independence of the annihilation $e^+e^-\to f\bar f$, we expect that 
such discovery should be more probable in that reaction. However, in this 
scenario, the usefulness of $e^+e^-\to W^+W^-$ would be not only to confirm the 
existence of the $Z_2$ but, especially important, to `probe' the 
$Z$-$Z^\prime$ mixing angle with the high accuracy indicated in Fig.~2. As 
previously mentioned, the highest sensitivity on $\phi$ would be obtained at 
CM energy $\displaystyle\sqrt s=M_{Z_2}\pm\frac{\Gamma_{Z_2}}{2}$, and 
quantitatively we represent it in Fig.~3, where the thick and thin solid lines 
correspond to $\sigma^L$ and $\sigma^R$ respectively, both for variable 
$\cos\beta$ and for the specific models $\psi$, $\chi$ and $\eta$. In 
particular, once again Fig.~3 shows also the complementary role of the two 
possible initial beam longitudinal polarizations. Actually, the tiny 
`around-resonance' $\phi$-values in Fig.~3 are more illustrative than really 
quantitative, because in this case the practical analysis should be 
supplemented by radiative corrections in that energy range, which would 
require separate consideration. For the `off-resonance' case 
$\sqrt s\ll M_{Z_2}$ this problem should be less important, because the 
$\phi$-values probed there correspond to relative amplitude deviations from 
the SM of order $0.1$, probably much larger than the effect of electroweak 
corrections \cite{lahanas}. 
\par
The analogue of Fig.~2 for the NLC500 is depicted in Fig.~4 for the case 
$\sqrt s=1\hskip 2pt TeV$ and $L_{int}=200\hskip 2pt fb^{-1}$. The bounds from 
process (\ref{proc}) at this energy are well consistent with those in Fig.~2 
through the anticipated approximate scaling law $\sqrt{L_{int}s}$ for the 
sensitivity $\cal S$. In Figs.~5 to 8 we 
report the results for the models $\eta$ and $\chi$, analogous to Figs.~2 and 
4. The general features pointed out above for the $\psi$ model also hold for 
these other cases, and the corresponding sensitivities can be directly read 
from the figures.
\par  
In conclusion, we have discussed the possibility of probing $Z$-$Z^\prime$ 
mixing at future $e^+e^-$ linear colliders, {\it via} the measurement of 
$e^+e^-\to W^+W^-$ cross sections with longitudinally polarized beams. While 
the corresponding bounds at the NLC500 are found to complement 
current ones, they rather rapidly improve at the higher energies. Also, such 
bounds have been compared to the ones generally obtainable from the 
consideration of the nondiagonal entries of the $Z$-$Z^\prime$ mass-matrix, 
as well as with the `superstring inspired' extended gauge models.
Moreover, if the $Z^\prime$ was discovered, presumably in $e^+e^-\to f\bar f$, 
measurements of $e^+e^-\to W^+W^-$ for energy around the $Z^\prime$ peak would 
provide either, perhaps, a `direct' measurement of ${Z-Z^\prime}$ mixing or,  
in any case, a strong limit on such effects. 
These features of $e^+e^-\to W^+W^-$ seem particularly useful in the 
case of the specific extended gauge models considered here, and the 
benefits of initial beams polarization in this kind of analysis are clearly 
manifest.
\vspace{2.0cm}
\section*{Acknowledgements}
\noindent AAP acknowledges the support and the hospitality of INFN-Sezione di 
Trieste and of the International Centre for Theoretical Physics, Trieste.
The work of NP has been supported in part by the Human Capital and Mobility 
Programme, EEC Contract ERBCHRXCT930132.

\newpage
\section*{Figure captions}
\begin{description}

\item{\bf Fig.~1} Upper bounds (95\% CL) on non-standard couplings
($\delta_\gamma, \delta_Z$) from $e^+e^-\to W^+W^-$
with longitudinally polarized electrons at $\sqrt s=0.5\hskip 2pt TeV$ 
and integrated luminosity $L_{int}= 50\hskip 2pt fb^{-1}$.
$\sigma^L$, $\sigma^R$ and $\sigma^{unpol}$ refer to allowed regions 
obtained from polarized cross sections with degrees of polarization 
$P_L=-0.9;\ +0.9;\ 0$, respectively. The straight dotted lines represent 
relation (\ref{relation}) for specific models: $\psi$, $\eta$, and $\chi$.
 
\item{\bf Fig.~2} Allowed domains (95\% C.L.) on ($\phi, M_{Z_2}$)
for the $\psi$ model. The thick solid contour corresponds to the region 
obtained at the NLC500 from $e^+e^-\to W^+W^-$. 
Also, the current limit on $M_2$ and the one expected from $e^+e^-\to l^+l^-$ 
at $\sqrt s=0.5\hskip 2pt TeV$ are indicated. The dotted lines correspond to 
the constraints derived from the $Z-Z^\prime$ mass-matrix (\ref{phi}) with 
$\Delta M=0.2\hskip 2pt GeV$. The thin solid contour shows the constraint 
for the `superstring' model case, Eq.~(\ref{models}) for $\sigma=0$ and 
$\sigma=\infty$. 

\item{\bf Fig.~3} Upper limits (95\% C.L.) for $\phi$ {\it vs.} the
$E_6$ model parameter $\cos\beta$ from  $W$ pair production
at ${\displaystyle M_{Z_2}=\sqrt s \pm \frac{\Gamma_{Z_2}}{2}}$
and with $P_L=-0.9$ (thick solid line) and $P_L=+0.9$ (thin solid line).

\item{\bf Fig.~4} Same as Fig.~2, for the NLC1000 with 
$\sqrt s=1\hskip 2pt TeV$ and $L_{int}=200\hskip 2pt fb^{-1}$. .

\item{\bf Fig.~5} Same as Fig.~2 for the $\eta$ model (NLC500).

\item{\bf Fig.~6} Same as Fig.~2 for $\eta$ model (NLC1000).

\item{\bf Fig.~7} Same as Fig.~2 for $\chi$ model (NLC500).

\item{\bf Fig.~8} Same as Fig.~2 for $\chi$ model (NLC1000).

\end{description}

\vfill\eject
\end{document}